\documentclass[sigconf]{acmart}
\AtBeginDocument{%
  }

\copyrightyear{2025}
\acmYear{2025}
\setcopyright{cc}
\setcctype{by-nc-nd}
\acmConference[SIGIR '25]{Proceedings of the 48th International ACM SIGIR Conference on Research and Development in Information Retrieval}{July 13--18, 2025}{Padua, Italy}
\acmBooktitle{Proceedings of the 48th International ACM SIGIR Conference on Research and Development in Information Retrieval (SIGIR '25), July 13--18, 2025, Padua, Italy}\acmDOI{10.1145/3726302.3729995}
\acmISBN{979-8-4007-1592-1/2025/07}

\usepackage{appendix}
\usepackage{multirow} 
\usepackage{algorithm}
\usepackage{subcaption}
\usepackage{algorithmic}
\usepackage{enumitem}
\DeclareMathOperator*{\argmax}{arg\,max} 

\begin{document}

\title{Hierarchical Tree Search-based User Lifelong Behavior Modeling on Large
Language Model}

\author{Yu Xia}
\authornote{Both authors contributed equally to this research.}
\orcid{0009-0002-4128-6968}
\affiliation{%
  \institution{University of Chinese Academy of Sciences}
  \city{Beijing}
  \country{China}
}
\email{xiayu24@mails.ucas.ac.cn}

\author{Rui Zhong}
\authornotemark[1]
\affiliation{%
  \institution{Kuaishou Technology}
  \city{Beijing}
  \country{China}
}
\email{zhongrui@kuaishou.com}

\author{Hao Gu}
\affiliation{%
  \institution{Kuaishou Technology}
   \city{Beijing}
  \country{China}
}
\email{guhao@kuaishou.com}

\author{Wei Yang}
\affiliation{%
 \institution{Kuaishou Technology}
  \city{Beijing}
  \country{China}
}
\email{yangwei08@kuaishou.com}

\author{Chi Lu}
\affiliation{%
  \institution{Kuaishou Technology}
  \city{Beijing}
  \country{China}
}
\email{luchi@kuaishou.com}

\author{Peng Jiang}
\affiliation{%
  \institution{Kuaishou Technology}
  \city{Beijing}
  \country{China}
}
\email{jiangpeng@kuaishou.com}

\author{Kun	Gai}
\affiliation{%
  \institution{Unaffiliated}
  \city{Beijing}
  \country{China}
}
\email{gai.kun@qq.com}

\renewcommand{\shortauthors}{Yu Xia et al.}

\begin{abstract}
Large Language Models (LLMs) have garnered significant attention in Recommendation Systems (RS) due to their extensive world knowledge and robust reasoning capabilities. However, a critical challenge lies in enabling LLMs to effectively comprehend and extract insights from massive user behaviors. Current approaches that directly leverage LLMs for user interest learning face limitations in handling long sequential behaviors, effectively extracting interest, and applying interest in practical scenarios. To address these issues, we propose a \textbf{\underline{Hi}}erarchical \textbf{\underline{T}}ree Search-based User \textbf{\underline{L}}ifelong \textbf{\underline{B}}ehavior \textbf{\underline{M}}odeling framework (\textbf{HiT-LBM}). HiT-LBM integrates Chunked User Behavior Extraction (CUBE) and Hierarchical Tree Search for Interest (HTS) to capture diverse interests and interest evolution of user. CUBE divides user lifelong behaviors into multiple chunks and learns the interest and interest evolution within each chunk in a cascading manner. HTS generates candidate interests through hierarchical expansion and searches for the optimal interest with \textbf{process rating model} to ensure information gain for each behavior chunk. Additionally, we design Temporal-Ware Interest Fusion (TIF) to integrate interests from multiple behavior chunks, constructing a comprehensive representation of user lifelong interests. The representation can be embedded into any recommendation model to enhance performance. Extensive experiments demonstrate the effectiveness of our approach, showing that it surpasses state-of-the-art methods.
\end{abstract}

\ccsdesc[500]{Information systems~Recommender systems}

\keywords{Large Language Models; Recommender Systems; User Behavior Modeling}

\maketitle

\section{Introduction}
Recommendation systems (RS) have become ubiquitous in online services, significantly enhancing user experiences in areas such as online shopping \cite{w1}, movie discovery \cite{w2} and music recommendation \cite{w3}. Recently, LLMs have achieved remarkable breakthroughs, demonstrating exceptional capabilities in comprehension and reasoning. This has spurred the rapid development of LLM-enhanced recommendation systems as a promising research direction.

User behavior modeling \cite{mamba4rec, bert4rec, llmrec,llmrec_4} is important in recommendation tasks, as user behavior encapsulates rich information about user interests. Traditional recommendation models such as DIN \cite{din} and DIEN \cite{dien} often rely on ID features for user behavior modeling, which lacks an understanding of semantic information of user and item, presenting certain limitations in increasingly data-rich recommendation scenarios. Moreover, traditional recommendation models only utilize a portion of the user behavior for modeling, neglecting the completeness of user behavior. Although SIM \cite{sim} addresses this to some extent by selecting top-$K$ behaviors based on the similarity of all historical behaviors, it still does not fully resolve the issue. LLMs \cite{gpt4, llama, qwen, deepseek} with extensive world knowledge and powerful reasoning capabilities, offer a promising solution to this problem. However, leveraging LLMs for user behavior modeling presents two significant challenges due to their inherent limitations: (1) Restricted by context length, LLMs cannot directly extract useful information from the textual context of long user behavior sequences. Furthermore, some studies \cite{context,hu2024can,li2023loogle} have found that even when the context length does not reach the LLMs' limit, LLMs tend to focus unevenly on the input, favoring the beginning and end parts, which can lead to biases when modeling long user behaviors with LLMs. Given the vast amount of user behavior in RS, LLMs cannot directly and effectively model lifelong user behavior. (2) Over time, user interests are constantly evolving \cite{dien}. When modeling user interests through LLMs, each input behavior is treated equally, making it difficult to understand the dynamic changes in user interests from a temporal perspective. However, understanding the trends in user interest changes is crucial for RS. 

To address the aforementioned challenges, inspired by the multi-turn dialogue of LLMs, we propose a Hierarchical Tree Search-based User Lifelong Behavior Modeling framework (HiT-LBM). The framework primarily consists of the following three modules:  (1) Chunked User Behavior Extraction (CUBE),  (2) Hierarchical Tree Search for Interests (HTS),  (3) Temporal-Ware Interest Fusion (TIF).

Given the limitations of LLMs in processing long texts, CUBE divides users' lifelong historical behaviors into appropriately sized behavior chunks. This allows the LLMs to learn user interests in a cascading manner across adjacent chunks, while also capturing changes in user interests by comparing the interests derived from the previous behavior chunk with those from the current. In the process of learning interest within behavior chunks, the presence of a cascading approach can easily lead to error accumulation or quality degradation \cite{cascaded}. Specifically, if the quality of the generated interest from the previous behavior chunk is poor, the low-quality result is passed as input to the next behavior chunk, causing the output quality of subsequent interest learning to be affected and potentially further deteriorated. To ensure the quality of interest generation for each behavior chunk, HTS designs two process ratinging models: Sequence Rating Model (SRM) and Point Rating Model (PRM), to evaluate the continuity and validity of the current interest, respectively. Based on these ratings, we design an interest hierarchy tree search. For each behavior chunk, multiple candidate interests are generated through hierarchical expansion (Best-of-N mode). These candidates are then scored by SRM and PRM, and the optimal interest for the current behavior chunk is determined through a search guided by these ratings. By employing the hierarchical tree search method, we not only mitigate the quality degradation issue inherent in the cascading approach but also reduce the impact of hallucination problems associated with LLMs themselves. Given the inherently sequential nature of user behaviors, we transform the textual knowledge from LLMs into compact representations and perform temporal fusion to aid recommendation tasks. First, the knowledge extracted from the LLM is encoded into dense representations. Then, we design a temporal-enhanced fusion module to integrate user interests from different chunks, resulting in a user lifelong interest representation. Finally, this long-term fused user interest representation can be incorporated into any recommendation model for prediction, thereby enhancing performance.

Our contributions can be summarized as follows:
\begin{itemize}[leftmargin=*]
\item We propose a straightforward and efficient Hierarchical Tree Search-based User Lifelong Behavior Modeling framework (HiT-LBM). This framework not only tackles the issue of lengthy text in user interest learning with LLMs but also enhances the quality of generating interests and interest changes with LLMs. 
\item The proposed HiT-LBM comprises three core components:  Chunked User Behavior Extraction, Hierarchical Tree Search and Temporal-Ware Interest Fusion. HiT-LBM can discover user's optimal interest path in a cascading paradigm, temporally fuse interest features from the optimal path to assist in recommendation.
\item We conduct comprehensive experiments on two public datasets and online industrial experiments to validate the effectiveness of HiT-LBM. Additionally, a series of ablation studies and analyses confirm that HiT-LBM can effectively address the limitations of long behavior sequences on LLM capabilities, enhance the quality of interest modeling by LLMs, and better assist recommendations.
\end{itemize} 

\section{Related Work}
\subsection{Traditional Recommendation Models}
Research on traditional recommendation models primarily focuses on two categories: feature interaction modeling and user behavior sequence modeling. Feature interaction models aim to capture relationship between features through mechanisms such as Factorization Machine \cite{deepfm,xdeepfm}, Cross Network \cite{dcn, dcnv1}, and attention mechanism \cite{attention}. For instance, DeepFM \cite{deepfm} and xDeepFM \cite{xdeepfm} integrate Factorization Machines with deep neural networks to model both low-order and high-order feature interactions simultaneously. DCN \cite{dcnv1} and DCNv2 \cite{dcn} explicitly learn feature interactions through cross networks, utilizing multi-layer structures to capture interactions of different orders. Additionally, AutoInt \cite{autoint} and FiBiNet \cite{fibinet} introduce attention-based feature interaction methods, enhancing model performance while providing feature importance weights. The second category focuses on user behavior sequence modeling, extracting useful information from users' historical behavior sequences. Since behavior sequences are typically represented as time-ordered ID features, researchers have proposed various sequence modeling structures, such as attention mechanisms \cite{attention} and GRU \cite{gru}, to capture dynamic patterns. For example, DIN \cite{din} uses attention mechanisms to dynamically model the relationship between target items and users' historical behaviors, while DIEN \cite{dien} further introduces a GRU-based interest evolution network combined with attention mechanisms to depict the evolution of user interests. However, all these models overlook the powerful capabilities of large language models, which consequently limits their performance.

\begin{figure*}[h]
\includegraphics[width=0.95\textwidth]{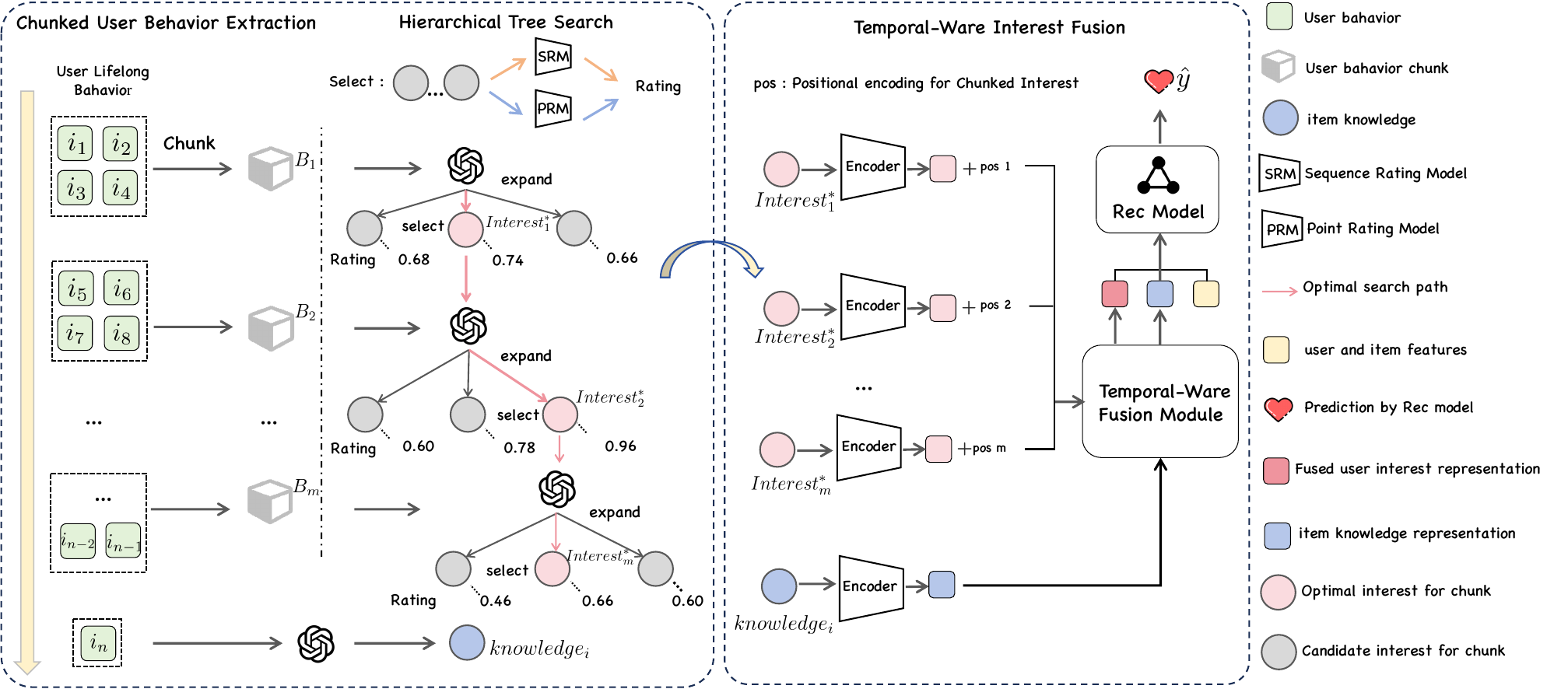}
\caption{Overview of the HiT-LBM framework. }
\label{overview}
\end{figure*}
\subsection{LLM for Recommendation }
LLM recommendation \cite{llmrec_1, llmrec_2,llmrec_3} is categorized into two paradigms: LLM as a recommender (LLM as Rec) \cite{tallrec, collm, binllm, rella} and LLM-enhanced recommendation (LLM enhance Rec) \cite{kar, liber}. For LLM as Rec, LLMs are employed as recommenders through zero-shot, few-shot, and fine-tuning strategies. For instance, TALLRec \cite{tallrec} leverages user interaction data for sequential recommendation by constructing a dataset from real user feedback and fine-tuning LLaMA-7B using LoRA. Similarly, CoLLM and BinLLM \cite{collm, binllm} enhance the recommendation performance of LLMs by incorporating ID feature during the fine-tuning process. To address the challenge of long-sequence recommendation, Rella \cite{rella} proposes a retrieval-based approach, where the $K$ most relevant items to the target item are retrieved and used as input for LLM fine-tuning. 

In the LLM enhance Rec paradigm, the capabilities of LLMs are harnessed to improve the performance of recommendation systems. For example, KAR \cite{kar} utilizes partial user behavior sequences and target items to extract user preferences and item knowledge via GPT-3.5, thereby facilitating auxiliary recommendation. Meanwhile, LiBER \cite{liber} adopts a partitioning strategy to generate multiple user summaries, which are subsequently aggregated into a unified user representation to enhance recommendation accuracy.

While these methods have demonstrated notable progress, they have yet to effectively resolve the critical challenge of comprehending long-sequence user behaviors within the context of user lifelong behavior modeling. Our proposed HiT-LBM framework addresses this limitation by fully harnessing the capabilities of LLMs, enabling efficient and effective lifelong user behavior modeling.

\section{Methodology}
\subsection{Overview}

To better leverage LLMs for lifelong user behavior modeling, we propose a Hierarchical Tree Search-based lifelong user behavior modeling framework, HiT-LBM. Figure \ref{overview} illustrates the overall architecture of our proposed HiT-LBM. This framework is model-agnostic and primarily consists of the following three modules:
\begin{itemize}[leftmargin=*]
\item[$\bullet$] \textbf{Chunked User Behavior Extraction (CUBE)}. We partition the user's lifelong historical behaviors into distinct behavior chunks of appropriate length, enabling the LLM to learn the user's interests in a cascading manner between adjacent chunks. 

\item[$\bullet$] \textbf{Hierarchical Tree Search for Interests(HTS)}. We design two process rating models—Sequence Rating Model (SRM) and Point Rating Model (PRM), to score the continuity and validity of current interests, respectively. Based on these scores, we devise an interest hierarchical tree search to identify the optimal interest for each behavior chunk. 

\item[$\bullet$] \textbf{Temporal-Ware Interest Fusion (TIF)}. Given the inherently sequential nature of user behaviors, we transform the textual knowledge from the LLM into compact representations and perform temporal fusion to assist recommendation tasks. First, the knowledge extracted from the LLM is encoded into dense representations. Then, we design a Temporal-Ware Fusion module to integrate user interests from different chunks, resulting in a user lifelong interest representation. Finally, the long-term fused user interest representation can be incorporated into any recommendation model for prediction, thereby enhancing performance.
\end{itemize} 

\subsection{Chunked User Bahavior Extraction}
\begin{figure*}[t]
\includegraphics[width=0.99\textwidth]{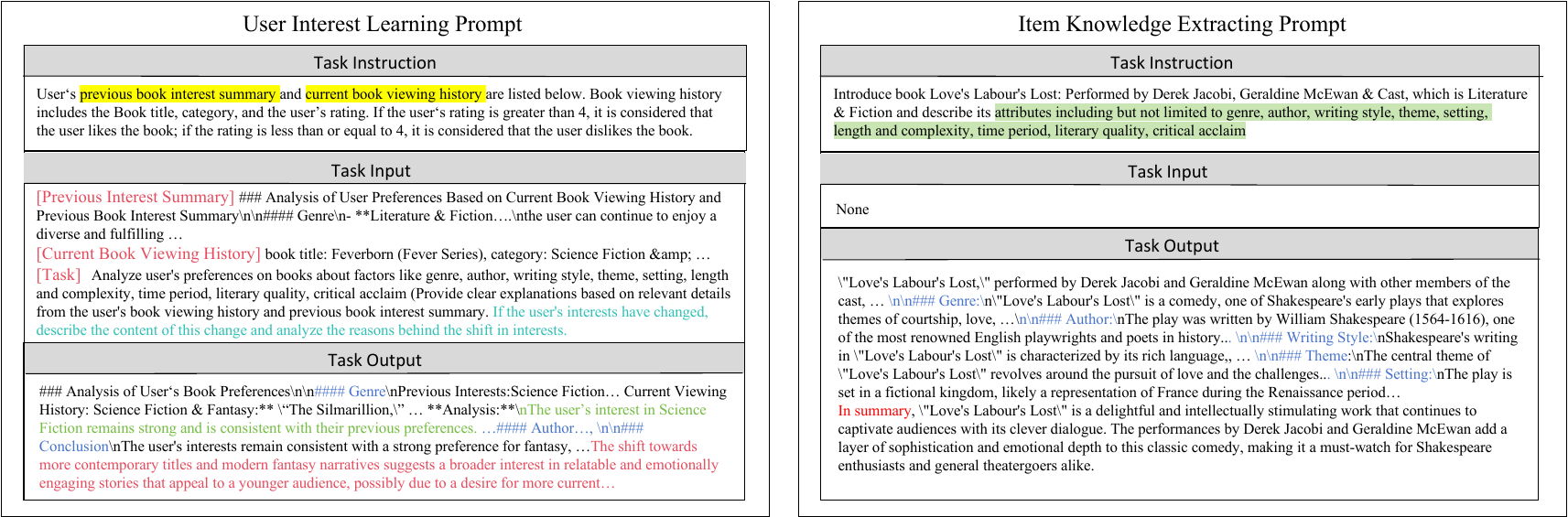}
\caption{Example prompts for User Interest Learning and Item Knowledge Extraction, denoted as $prompt^{inst}$ and $prompt^{item}$ respectively.}
\label{pprompt}
\end{figure*}

Due to the limitations of input text length, LLMs can not directly model lifelong user behaviors. Additionally, user interests are constantly evolving, making it challenging for LLMs to capture the dynamic changes in interests from lifelong user behaviors. Inspired by multi-turn dialogue, we can transform lifelong user behavior modeling into a multi-turn cascaded modeling approach. Specifically, we divide the user's long sequence of behaviors into multiple chunks and then jointly model these chunks. To achieve this, we select an appropriate chunk length \( L \) based on the LLM's context window size and partition each user's historical interactions into corresponding chunks. 
Consequently, we can obtain the division of the user's behavior chunks as follows:
\begin{equation}
    B = [B_1, B_2, \dots, B_m]
\end{equation}

Subsequently, we leverage the extensive world knowledge and robust reasoning capabilities of LLMs to analyze user interests, encompassing both current interests and their evolution. We have meticulously crafted a prompt instruction template, denoted as $prompt^{inst}$, for summarizing user interests and uncovering shifts in those interests, as illustrated in \textit{User Interest Learning Prompt} of Figure \ref{pprompt} . The $prompt^{inst}$ incorporates the user's current behavior chunk and tailored instruction for interest learning that align with the specific dataset, enabling the LLM to comprehensively capture user interests from multiple perspectives. Moreover, although we partition user behavior into distinct chunks, these chunks are not isolated but are intricately interconnected. Specifically, when generating the interests for the current behavior chunk, it is essential to consider not only the interactive behaviors within the current chunk but also the interest derived from previous chunk. This consideration forms the cornerstone for learning interest transitions. Therefore, we propose a cascading paradigm that integrates past user interest summaries and instructions for detecting interest shifts within $prompt^{inst}$. This approach ensures that the LLM, while thoroughly learning user interests, places greater emphasis on the evolution of new interests rather than reiterating old ones, aligning with human nature. We generate user interests and their changes through prompt-based learning as follows:
\begin{equation}
    Inst_i = 
\begin{cases} 
    LLM\left(prompt^{inst}, Inst_{i-1}, B_i\right), & \text{if } i > 1, \\
    LLM\left(prompt^{inst}, B_i\right),              & \text{otherwise}.
\end{cases}
\end{equation}
Here, \( {Inst}_i \) represents the user's interest in the $i^{th}$ behavior chunk.
\subsection{Hierarchical Tree Search for Interests}
To mitigate the issue of quality degradation in the cascading paradigm and ensure the information gain of each behavior chunk's interest during the cascading process, inspired by the process reward models in LLMs, we have designed a hierarchical tree-based interest search module. This module specifically includes two components: (1) \textbf{Process Rating Model}. We have devised two process rating models— Sequence Rating Model (SRM) and Point Rating Model (PRM). SRM and PRM are utilized to score the continuity and validity of the interests generated by the LLM, thereby assessing the quality of the interests produced by the LLM. (2) \textbf{Hierarchical Tree Search}. We formalize the modeling of user chunked behaviors into an interest tree. Starting from the root node, each layer's node represents the interests of the corresponding behavior chunk. In conjunction with the process scoring models, we can perform a greedy search on the interest tree to identify the optimal path for interest modeling.

\subsubsection{Process Rating Model}
\paragraph{Training Data Construction}
\label{data}
\begin{figure*}
\includegraphics[width=1\textwidth]{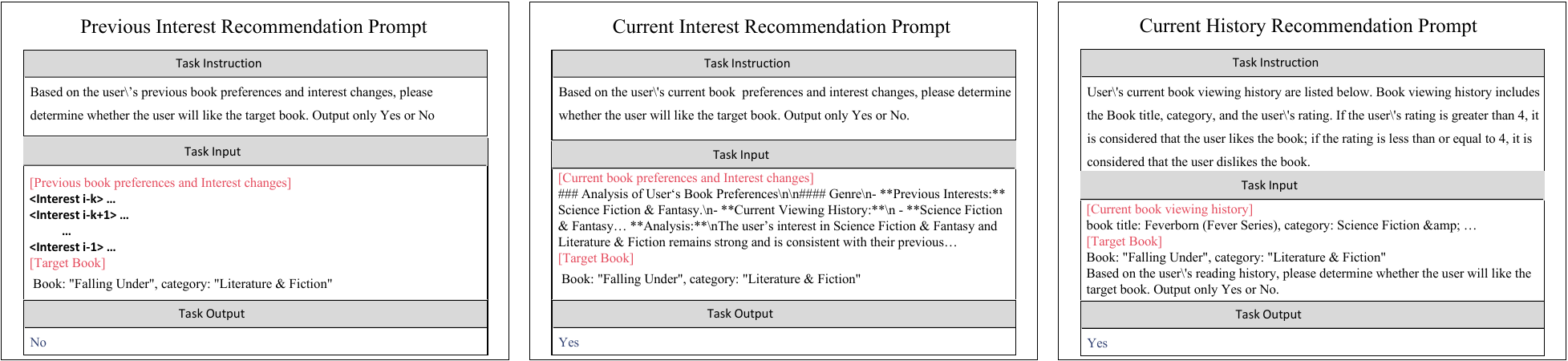}
\caption{The prompts we construct for the recommendation task incorporate the user's previous    interest ($prompt^{seq}$), current interest ($prompt^{point}$), and historical behavior ($prompt^{hist}$) respectively.}
\label{pprompt_rating}
\end{figure*}

To evaluate the quality of interests generated by the LLM for a particular behavior chunk, we utilize the items from the subsequent behavior chunk as target items. For a more comprehensive assessment, we randomly select $N$ positive samples and $N$ negative samples from the next behavior chunk to form the evaluation set, as detailed below:
\begin{equation}
S_t = \{(x_i, y_i)\}_{i=1}^{2N} \quad \text{where} \quad (x_i,y_i)\in B_{t+1}
\end{equation}
Here, $S_t$ represents the evaluation set for the $t^{th}$ chunk, $x_i$ denotes the item selected from the $t+1^{th}$ chunk, and $y_i$ denotes the label of $x_i$.

To assess the quality of interests from multiple perspectives, we evaluate the continuity and validity of interests by comparing the gains from previous interests to current interests, as well as the gains from excluding and including the current interest, thereby constructing the labels. To achieve this, we employ the LLM to introduce different user features for predictive discrimination on the evaluation set $S_t$. We construct three prompt instruction templates—$prompt^{seq}$, $prompt^{point}$, and $prompt^{hist}$, corresponding to the introduction of previous interests, the current interest, and the historical interaction items of the current behavior chunk, respectively, as illustrated in Figure \ref{pprompt_rating}. To prevent the prompt input from becoming excessively long, we only introduce the previous $K$ interests. Specifically, we instruct the LLM to reference the previous $K$ interests, the current interest, and the current behavior chunk to predict $S_t$, obtaining the probability corresponding to the LLM's reply with the "Yes" token. The details are as follows:
\begin{equation}
\begin{aligned}
    Y^{seq} &= \{P(\text{'Yes'} \mid Prompt^{seq}, [Inst_{t-K}, \dots, Inst_{t-1}], x_i)\}_{i=1}^{2N}, \\
    Y^{point} &= \{P(\text{'Yes'} \mid Prompt^{point}, Inst_t, x_i)\}_{i=1}^{2N}, \\
    Y^{hist} &= \{P(\text{'Yes'} \mid Prompt^{hist}, B_t, x_i)\}_{i=1}^{2N}.
\end{aligned}
\end{equation}
Here, $Inst_j$ represents the user interest of the $j^{th}$ behavior chunk, $B_t$ denotes the current behavior chunk, and $Y^{seq}$, $Y^{point}$, $Y^{hist}$ respectively indicate the probabilities corresponding to the prediction of "Yes" for each item in the evaluation set $S_t$ when different user features are introduced.

Subsequently, we calculate the Area Under the Curve (AUC) within the evaluation set for each interest chunk as follows:
\begin{equation}
     AUC^{seq},AUC^{point},AUC^{hist}= \text{AUC}(Y^{seq},Y^{point},Y^{hist},S_t)
\end{equation}
We posit that if the predictive AUC when referencing the current interest exceeds that of referencing the previous $K$ interests, then the current interest generated through the cascading paradigm not only continues the previous interests but also contributes information gain. We label the current interest as a positive sample; otherwise, it is labeled as a negative sample and added to the continuity scoring dataset $D_{cont}$. If the predictive AUC when referencing the current interest surpasses that of referencing the current behavior chunk alone, it indicates that the current interest effectively represents the user's present interests, and we label it as a positive sample; otherwise, it is labeled as a negative sample and added to the effectiveness scoring dataset $D_{eff}$. 
\paragraph{Model Training}
We employ a straightforward yet efficient binary classification method to implement the Process Rating Model. Initially, we utilize an encoder to obtain the continuity dense vector and the effectiveness dense vector of the current behavior chunk's interests. We can derive the following:
\begin{equation}
\begin{aligned}
{emb}_{cont} &= \mathcal{}{Aggr}\left({Encoder}\left({Inst_{t-K\ :\ t-1}, Inst_t}\right)\right), \\
{emb}_{eff} &= \mathcal{}{Aggr}\left({Encoder}\left({Inst_t}\right)\right).
\end{aligned}
\end{equation}
Here, $Inst_{t-K\ :\ t-1}$ represents the set of interests from the user's $(t-K)^{th}$ chunk to the $(t-1)^{th}$ chunk. $Aggr$ denotes the aggregation function.

Subsequently, we can employ a Multilayer Perceptron (MLP) to transform $emb_{cont}$ and $emb_{eff}$ into the final scores for interest continuity and effectiveness, respectively. Two distinct MLP along with the encoder constitute SRM and PRM, respectively. Following this, SPM and PRM will be trained using binary cross-entropy loss on the dataset $D_{conf}$ and $D_{eff}$. SRM and PRM will play a pivotal role in the hierarchical tree search.

\begin{figure}[h]
\includegraphics[width=0.45\textwidth]{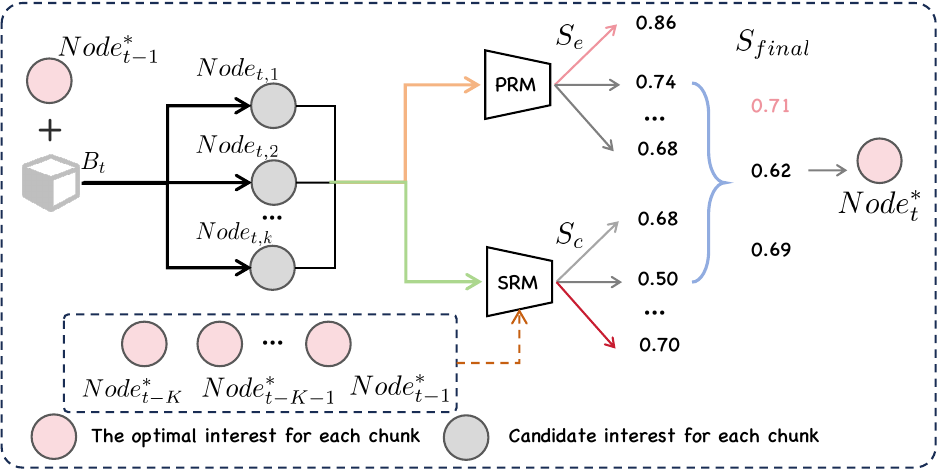}
\caption{The detailed workflow of how the Process Rating  Model functions in hierarchical tree search. }
\label{search}
\end{figure}

\subsubsection{Hierarchical Tree Search}

\begin{algorithm}
\caption{Hierarchical Tree Search for Interests}
\label{alg:hierarchical_tree_search}
\renewcommand{\algorithmicrequire}{\textbf{Input:}}
\renewcommand{\algorithmicensure}{\textbf{Output:}}
\begin{algorithmic}
\REQUIRE { User behavior chunks $B = \{B_1, B_2, \dots, B_T\}$,Sequence Rating Model $SPM$, Point Rating Model $PRM$, Number of expansions per node $N$, Maximum tree depth $T$, $SPM$ Reference Number of Interests $\mathcal{K}$
    }
\ENSURE {Hierarchical interest tree: $\mathcal{T}$}

\STATE Initial: root node $Node_0$, tree $\mathcal{T} \gets \{Node_0\}$, $\mathcal{P} \gets \emptyset$

\FOR{each layer $i \in \{1, 2, \dots, T\}$}
    \STATE \textbf{Step 1: Node Expansion} \\
    \quad Select parent node $Node_{i-1}^*$ in layer $i-1$: \\
    \quad a. Generate $N$ child nodes using LLM's Best-of-N sampling:
    \[
    Children_{i-1} = \{Node_{i,k}\}_{k=1}^N, \quad Node_{i,k} \sim \text{LLM}(Node_{i-1}^*, B_i)
    \]
    \quad b. Add $Children_{i-1}$ to $\mathcal{T}$.

    \STATE \textbf{Step 2: Node Rating} \\
    \FOR{each child node $Node_{i,k} \in Children_{i-1}$}
    \STATE a. Compute continuity score $S_c$ using $SPM$:
    \[
    S_c(Node_{i,k}) = SPM([Node_{i-\mathcal{K}}^*,Node_{i-\mathcal{K}+1}^*,Node_{i-1}^*], Node_{i,k})
    \]
    b. Compute effectiveness score $S_e$ using $PRM$:
    \[
    S_e(Node_{i,k}) = PRM(Node_{i,k})
    \]
    c. Compute final score $S_{final}$:
    \[
    S_{final}(Node_{i,k}) = \alpha S_c(Node_{i,k}) + (1 - \alpha) S_e(Node_{i,k})
    \]
    \ENDFOR
    \STATE \textbf{Step 3: Node Selection} \\
    \quad Select the child node with the highest $S_{final}$:
    \[
{Node}_i^* = \argmax_{\text{Node}_{i,k} \in \text{Children}_{i-1}} S_{final}({Node}_{i,k})
\]
    \[
     \mathcal{P} = \mathcal{P} \cup \{{Node}_i^*\}
    \]
    \quad Set $Node_i^*$ as the new parent node for the layer $i$.
\ENDFOR

\RETURN $\mathcal{T}$
\end{algorithmic}
\end{algorithm}

To enhance the quality of chunked user behavior modeling and alleviate the issue of interest quality degradation under the cascading paradigm, we introduce hierarchical tree search into user interest learning, as illustrated in the algorithm \ref{alg:hierarchical_tree_search}. We abstract the chunked user behavior modeling into an interest tree, where the $i^{th}$ layer of the tree represents the $i^{th}$ behavior chunk $B_i$. Each node $Node_{i,k}$ at the $i^{th}$ layer denotes the interest of the $i^{th}$ behavior chunk $B_i$. For each user, we begin by initializing a root node and expanding from there. The expansion of each node depends on its parent node and the corresponding behavior chunk at the current layer. For each expanded node, we perform $N$ expansions, utilizing the LLM's Best-of-N mode to sample multiple times and generate several child nodes $Node_{j,k}$. Subsequently, as shown in the Figure \ref{search} , we use the previously trained SRM and PRM to score each child node $Node_{j,k}$, obtaining the continuity score $S_c$ and the effectiveness score $S_e$ for each node. We calculate the final score $S_{final}$ ("rating" in Figure \ref{overview}) for each child node through a weighted approach:
\begin{equation}   
    S_{final}(Node_{j,k}) = \alpha S_c(Node_{j,k}) + (1 - \alpha) S_e(Node_{j,k})
\end{equation}
According to $S_{final}$, the child nodes are sorted, and the child node with the highest $S_{final}$ is selected as the new expansion node for the next round of expansion, continuing until the depth of the tree reaches the maximum number of user chunks. The path composed of all expanded nodes represents the optimal search path:
\begin{equation}
    \mathcal{P} = \{Node_1^*, Node_2^*,\dots, Node_T^*\}
\end{equation}
Here, $Node_{i}^*$ denotes the expanded node at the $i^{th}$ layer, which represents the optimal interest of the $i^{th}$ behavior chunk.

We extract the corresponding user interests from the nodes of the optimal search path $P$ to serve as the final user interests:
\begin{equation}
     \mathcal{I} = \{Inst_1^*, Inst_2^*,\dots, Inst_T^*\}
\end{equation}

\subsection{Temporal-Ware Interest Fusion}
Incorporating user interests modeled by LLMs into recommendation systems presents new challenges for us. Given the inherently sequential nature of user behaviors, we design a Temporal-Ware Interest Fusion, which comprises three components: Interest Encoder, Temporal-Ware Fusion, and Interest Utilization.
\subsubsection{Interest Encoder}
To harness the semantic information of interests generated by LLMs, we devise an interest encoder that encodes the interest text output by LLMs. This interest encoder is a lightweight, compact language model capable of swiftly transforming text into vector representations. Through this method, we can obtain the dense vector representation $e_j$ of the chunked interest $Inst_j$:
\begin{equation}
    e_j = Encoder(Inst_j)
\end{equation}

\subsubsection{Temporal-Ware Fusion}
Given that nodes in the hierarchical tree possess varying qualities, we utilize the final scores from the search process to weight the interests. Taking into account the sequential nature of user behavior, we assign a positional encoding to the interest vector of each chunk, yielding:
\begin{equation}
e_j^* = e_j * S_{final}^j  + pos_j
\end{equation}
We employ a Masked Self-Attention layer to first integrate the dense representations of different chunks, ensuring that the representation at each position is only aware of the interest representations preceding it, thereby preserving the temporal sequence in the fusion. The detailed process is defined as follows:
\begin{equation}
{MSA}({R})={Softmax}\left(\frac{{RW}^Q({RW}^K)^T}{\sqrt{d}}+Mask\right){RW}^V 
\end{equation}
Here, $d$ represents the dimension of the interest vector. $W_Q$, $W_K$, and $W_V$ represent the weights for query, key, and value respectively, and $R$ is the concatenation of $(e_1^*, e_2^*, ..., e_T^*)$. $Mask$ is an upper triangular matrix where the upper triangle (excluding the diagonal) is entirely filled with 1, and the rest are 0. The MSA computes the weights for different chunks by controlling visibility to obtain a unified fused weighted representation $R^* (e_1^f, e_2^f, ..., e_T^f)$.

Subsequently, we aim to introduce the target item to enhance the fusion of chunked interests into the final representation. As shown in the \textit{item Knowledge Extracting Prompt} of Figure \ref{pprompt}, we design an item knowledge extraction prompt template $prompt^{item}$, through which we utilize the LLM to acquire knowledge about the target item and, similar to user interests, transform it into a dense vector $e^{item}$. Finally, a cross-attention layer($CrossAttn$) is employed to complete the interest enhancement fusion guided by item knowledge:
\begin{equation}
    e^{user}=\mathrm{Aggr}\left({Softmax}\left(\frac{{R^*W}^Q({e_i^{item} W}^K)^T}{\sqrt{d}}\right){e_i^{item} W}^V\right)
\end{equation}
Here, $e_i^{item}$ denotes the vector of the item to be predicted. After $CrossAttn$, final user interest representation $e^{user}$ is thus obtained.
\subsubsection{Interest Utilization}
The LLM-enhanced user interest representations and item knowledge obtained through fusion can be integrated into traditional recommendation models. Specifically, we incorporate them as side information within the recommendation model. We formalize this as follows:
\begin{equation}
    \hat{y} = f(r^{user},r^{item}, e^{user}, e^{item}; \theta)
\end{equation}
Here, $r^{user}$ and $r^{item}$ represent the user ID features and target item ID features of the traditional recommendation model, respectively, while $e^{user}$ and $e^{item}$ denote the user semantic features and item semantic features extracted by the LLM, respectively. $r^{user}$ is derived from the user's recent behaviors, whereas $e^{user}$ originates from the user's lifelong behaviors and takes into account the evolution of the user's interests. Introducing long-term user interests and their evolution into traditional recommendation models can better assist the models in completing recommendation tasks and enhance the user experience.

\section{Experiments}
In this section, we conduct extensive experiments on both public and industry datasets to answer the following questions:
\begin{itemize}[leftmargin=*]
    \item \textbf{RQ1:} What improvements can HiT-LBM bring to traditional models on CTR task?
    \item \textbf{RQ2:} How does HiT-LBM compared with other user behavior modeling baseline models?
    \item \textbf{RQ3:} Does HiT-LBM gain performance improvement when deployed online?
    \item \textbf{RQ4:} What is the influence of different components in HiT-LBM on recommendation performance?
    \item \textbf{RQ5:} What is the impact of the interest fusion module on recommendation performance?
    \item \textbf{RQ6:} Does the parameter scale of LLM have an impact on user interest generation?
    \item \textbf{RQ7:} What is the impact of different interest encoders on the performance of HiT-LBM?
\end{itemize}

\subsection{Experimental Setup}
\begin{table}[htbp]
\centering
\caption{Statistics of datasets used in this paper} 
\setlength{\tabcolsep}{3pt} 
\begin{tabular}{lcccccc}
\toprule
{Dataset} & {User} & {Item} & {Train} & {Test} & {Chunks} & {Tokens} \\
\midrule
Amazon Book & 18750 & 22851 & 1015861 & 216437 & 55 & 1973 \\
MovieLens-1M  & 4297  & 3689  & 633150 & 95471   & 47 & 2655 \\
\bottomrule
\end{tabular}
\label{tab:dataset_stats}
\end{table}
\subsubsection{Datasets}
Experiments are conducted for the following two public datasets and one industrial dataset. {{\bfseries MovieLens-1M\footnote{\url{https://movielens.org/}}}} is a movie review dataset. Following previous work \cite{dien,din}, we label samples with ratings above 3 as positive and the rest as negative. We filter out users with fewer than 50 reviews. The dataset is divided into a training set and a test set in a 9:1 ratio based on global timestamps. For process rating model, we construct one sample for each chunk of the training set, resulting in a total of 16,852 samples. {{\bfseries Amazon Book\footnote{\url{https://nijianmo.github.io/amazon/index.html}}}} is the "Books" category from the Amazon Review Dataset. The preprocessing procedure is similar to that of MovieLens-1M. Moreover, samples with a rating of 5 are considered positive, while the rest are deemed negative. For process rating model, there are a total of 35,405 samples. \textbf{Industrial Dataset} is collected from Kuaishou's advertising platform with hundreds of millions of users. Samples are constructed through sampling from impression logs. We chunk the user behaviors that occurred within Kuaishou (over one year) on a monthly basis. The statistics of the processed datasets are presented in Table \ref{tab:dataset_stats}. "Chunks" and "Tokens" respectively denote the maximum number of chunks after partitioning and the average number of tokens per chunk.

\begin{table*}
  \caption{Effectiveness Analysis of HiT-LBM Across Different Backbone CTR Models on MovieLens and Amazon-books Datasets.}
  \label{tab:effect}
  \begin{tabular}{cccccccccccccc}
    \toprule
    \multirow{3}{*}{\textbf{Models}}  &\multicolumn{6}{c}{\textbf{MovieLens-1M}} &\multicolumn{6}{c}{\textbf{Amazon-Book}} \\ 
    \cline{2-13}
    & \multicolumn{3}{c}{\textbf{AUC}}& \multicolumn{3}{c}{\textbf{LogLoss}}&\multicolumn{3}{c}{\textbf{AUC}}& \multicolumn{3}{c}{\textbf{LogLoss}} \\
    \cline{2-13}
    & \multicolumn{1}{c}{\textbf{Base}} &\multicolumn{1}{c}{\textbf{Our}}&\multicolumn{1}{c}{\textbf{Rel.Impr.}} &  \multicolumn{1}{c}{\textbf{Base}} &\multicolumn{1}{c}{\textbf{Our}}& \multicolumn{1}{c}{\textbf{Rel.Impr.}} & \multicolumn{1}{c}{\textbf{Base}} &\multicolumn{1}{c}{\textbf{Our}}& \multicolumn{1}{c}{\textbf{Rel.Impr.}} &\multicolumn{1}{c}{\textbf{Base}} &\multicolumn{1}{c}{\textbf{Our}} &\multicolumn{1}{c}{\textbf{Rel.Impr.}} \\
    \midrule
    DIEN & 0.7744 & \textbf{0.8064\textsuperscript{*}} & \textcolor{red!60}{$\uparrow$\,4.13\%} & 0.5611 & \textbf{0.5321\textsuperscript{*}} & \textcolor{blue}{$\downarrow$\,5.16\%} & 0.8220 & \textbf{0.8366\textsuperscript{*}} & \textbf{\textcolor{red!60}{$\uparrow$\,1.78\%}} & 0.5076 & \textbf{0.4920\textsuperscript{*}} & \textcolor{blue}{$\downarrow$\,3.07\%} \\
    DIN & 0.7761 & \textbf{0.8063\textsuperscript{*}} & \textcolor{red!60}{$\uparrow$\,3.89\%} & 0.5583 & \textbf{0.5301\textsuperscript{*}} & \textcolor{blue}{$\downarrow$\,5.05\%} & 0.8270 & \textbf{0.8385\textsuperscript{*}} & \textcolor{red!60}{$\uparrow$\,1.39\%} & 0.5019 & \textbf{0.4886\textsuperscript{*}} & \textcolor{blue}{$\downarrow$\,2.65\%} \\
    DCNv2 & 0.7747 & \textbf{0.8081\textsuperscript{*}} & \textcolor{red!60}{$\uparrow$\,4.31\%} & 0.5600 & \textbf{0.5307\textsuperscript{*}} & \textcolor{blue}{$\downarrow$\,5.23\%} & 0.8283 & \textbf{0.8364\textsuperscript{*}} & \textcolor{red!60}{$\uparrow$\,0.98\%} & 0.4987 & \textbf{0.4928\textsuperscript{*}} & \textcolor{blue}{$\downarrow$\,1.18\%} \\
    DeepFM & 0.7746 & \textbf{0.8063\textsuperscript{*}} & \textcolor{red!60}{$\uparrow$\,4.09\%} & 0.5608 & \textbf{0.5318\textsuperscript{*}} & \textcolor{blue}{$\downarrow$\,5.17\%} & 0.8281 & \textbf{0.8372\textsuperscript{*}} & \textcolor{red!60}{$\uparrow$\,1.10\%} & 0.4991 & \textbf{0.4924\textsuperscript{*}} & \textcolor{blue}{$\downarrow$\,1.34\%} \\
    xDeepFM & 0.7750 & \textbf{0.8029\textsuperscript{*}} & \textcolor{red!60}{$\uparrow$\,3.60\%} & 0.5599 & \textbf{0.5404\textsuperscript{*}} & \textcolor{blue}{$\downarrow$\,3.48\%} & 0.8280 & \textbf{0.8371\textsuperscript{*}} & \textcolor{red!60}{$\uparrow$\,1.10\%} & 0.5002 & \textbf{0.4870\textsuperscript{*}} & \textcolor{blue}{$\downarrow$\,2.64\%} \\
    AutoInt & 0.7745 & \textbf{0.8077\textsuperscript{*}} & \textcolor{red!60}{$\uparrow$\,4.29\%} & 0.5616 & \textbf{0.5325\textsuperscript{*}} & \textcolor{blue}{$\downarrow$\,5.18\%} & 0.8279 & \textbf{0.8404\textsuperscript{*}} & \textcolor{red!60}{$\uparrow$\,1.51\%} & 0.5000 & \textbf{0.4894\textsuperscript{*}} & \textcolor{blue}{$\downarrow$\,2.12\%} \\
    FiBiNet & 0.7749 & \textbf{0.8086\textsuperscript{*}} &  \textbf{\textcolor{red!60}{$\uparrow$\,4.35\%}} & 0.5600 & \textbf{0.5360\textsuperscript{*}} & \textcolor{blue}{$\downarrow$\,4.28\%} & 0.8276 & \textbf{0.8386\textsuperscript{*}} & \textcolor{red!60}{$\uparrow$\,1.33\%} & 0.5008 & \textbf{0.4872\textsuperscript{*}} & \textcolor{blue}{$\downarrow$\,2.72\%} \\
    FiGNN & 0.7744 & \textbf{0.8076\textsuperscript{*}} & \textcolor{red!60}{$\uparrow$\,4.29\%} & 0.5606 & \textbf{0.5310\textsuperscript{*}} & \textbf{\textcolor{blue}{$\downarrow$\,5.28\%}} & 0.8271 & \textbf{0.8401\textsuperscript{*}} & \textcolor{red!60}{$\uparrow$\,1.57\%} & 0.5015 & \textbf{0.4845\textsuperscript{*}} & \textbf{\textcolor{blue}{$\downarrow$\,3.39\%}} \\
    \bottomrule
    \multicolumn{13}{l}{* denotes statistically significant improvement (measured by t-test with p-value<0.001) over baselines.}
  \end{tabular}
\end{table*}

\subsubsection{Backbone Models}
Due to the model-agnostic nature of HiT-LBM, we select several classic CTR and CVR models to validate its effectiveness. \textbf{AutoInt} ~\cite{autoint} adopts a self-attentive neural network with residual connections to model the interactions explicitly.
\textbf{DeepFM} ~\cite{deepfm} use factorization machine to capture low-order and high-order feature interactions.  
\textbf{xDeepFM} ~\cite{xdeepfm} propose a novel Compressed Interaction Network, which aims to generate feature interactions in an explicit fashion and at the vector-wise level.  
\textbf{FiGNN} ~\cite{fignn} design a novel model Feature Interaction Graph Neural Networks to take advantage of the strong representative power of graphs.
\textbf{FiBiNet} ~\cite{fibinet} is an abbreviation for Feature Importance and Bilinear feature Interaction NETwork is proposed to dynamically learn the feature importance and fine-grained feature interactions.
\textbf{DCN} ~\cite{dcn} incorporates cross-network architecture to learn the bounded-degree feature interactions. 
\textbf{DIN} ~\cite{din} utilizes attention to model user interests dynamically with respect to a certain item. 
\textbf{DIEN} ~\cite{dien} introduces an interest evolving mechanism to capture the dynamic evolution of user interests over time. 

\subsubsection{Competitors}
We compared other user behavior modeling methods, including (1) traditional sequential recommendation models: DIN\cite{din} and SIM\cite{sim}, and (2) LLM-enhanced recommendation models: KAR\cite{kar}, TRSR\cite{trsr}, and LIBER\cite{liber}. To ensure a fair comparison, we standardized the data and features for all methods. Below is a brief introduction to these methods: \textbf{SIM} models the user by selecting the top-K behaviors based on the similarity retrieval of all the user's historical behaviors. \textbf{KAR} utilizes a portion of the user's behavior sequence and the target item to extract user preferences and item knowledge through GPT-3.5 for auxiliary recommendations. \textbf{LIBER} employs a partitioning method to extract multiple user summaries, which are then fused into a single user representation for enhanced recommendations. \textbf{TRSR} adopts a progressive approach to generate user preference summaries, using the most recent user interests as enhancement information.

\subsubsection{Parameter Configuration}
We set the chunk length $L$=50, the number of samples $N$=6 for each chunk, the number of interests $K$=1 referenced by the sequence rating model SPM, score fusion $\alpha$ = 0.5, and the number of expansions for each node in the interest tree to 10. For both the process scoring model and interest utilization, we use BGE\cite{bge} as the encoder. We employ Qwen2.5-7B\cite{qwen} as the interest generator. Other parameters, such as batch size and learning rate, are determined through grid search to achieve the best results. For fair comparisons, the parameters of the backbone model and the baselines are also
tuned to achieve their optimal performance.

\subsubsection{Evaluation Metrics}
We use AUC and LogLoss (binary crossentropy loss) to evaluate the performance of our proposed
model. A higher AUC value or a lower Logloss value, even by a small margin can be viewed as a significant improvement in CTR and CVR prediction performance.

\subsection{Offline Experimental Results}
\subsubsection{Improvement over different traditional Models (RQ1). }
We implement our proposed HiT-LBM method across multiple representative CTR (Click-Through Rate) and CVR (Conversion Rate) prediction models. Specifically, we integrate the final user and item representations obtained from the temporal-aware fusion module as side information into traditional recommendation models to observe its impact on various models. The results, as shown in Table \ref{tab:effect}, lead us to the following observations: (i) The application of HiT-LBM significantly enhances the performance of various baseline models, with an average improvement of 4.12\% on MovieLens-1M and 1.35\% on the Amazon Book. This validates that HiT-LBM effectively learns user interests and their evolution, liberating the constraints of long sequences on LLMs, thereby boosting the performance of recommendation models. (ii) As a model-agnostic framework, HiT-LBM can be applied to a wide range of recommendation models. By incorporating our framework, the selected representative models exhibit significant performance improvements in AUC and Logloss on both public datasets, demonstrating the versatility of HiT-LBM.

\subsubsection{Superiority: Improvement over Baselines (RQ2).}
\begin{table}
  \caption{Performance Comparison with baselines on MovieLens-1M and
Amazon-book Datasets. The best results are highlighted in bold, while the second-best results are indicated with underlining.}
  \label{tab:compare}
  \begin{tabular}{cccccc}
    \toprule
    \multirow{2}{*}{\textbf{Method}} & \multicolumn{2}{c}{\textbf{MovieLens-1M}} & \multicolumn{2}{c}{\textbf{Amazon-Book}} \\
    \cline{2-5}
    & \textbf{AUC} & \textbf{LogLoss} & \textbf{AUC} & \textbf{LogLoss} \\
    \midrule
    DIN & 0.7761 & 0.5583 & 0.8270 & 0.5019 \\
    SIM & 0.7853 & 0.5512 & 0.8314 & 0.4963 \\
    KAR & 0.7935 & 0.5435 & 0.8316 & 0.4959 \\
    TRSR & 0.7818 & 0.5547 & 0.8339 & \underline{0.4915} \\
    LIBER & \underline{0.7952} & \underline{0.5419} & \underline{0.8343} & 0.4918 \\
    \textbf{Ours} & \textbf{0.8063} & \textbf{0.5301} & \textbf{0.8385} & \textbf{0.4886} \\
    \bottomrule
  \end{tabular}
\end{table}
Next, we compare HiT-LBM with recent baseline methods for user behavior modeling, which include two categories: traditional sequential recommendation models such as DIN\cite{din} and SIM\cite{sim}, and LLM-enhanced recommendation models like KAR\cite{kar}, TRSR\cite{trsr}, and LIBER\cite{liber}. In the comparison of LLM-enhanced recommendation models, we uniformly use DIN as the backbone recommendation model. The results are shown in Table~\ref{tab:compare}, from which we can draw the following observations: (i) LLMs, with their superior text comprehension capabilities, achieve better performance in user behavior modeling compared to traditional ID-based recommendation models. (ii) In contrast to other LLM-enhanced recommendation methods, HiT-LBM can fully and effectively learn from vast amounts of user behavior, fully leveraging the comprehension capabilities of LLMs to enhance recommendation performance. (iii) Our model significantly outperforms other LLM-based baseline methods, achieving notable improvements in both AUC and Logloss metrics.

\subsection{Online Experimental Results}
\subsubsection{Experimental Setup}
To validate the effectiveness of HiT-LBM in real-world scenarios, we conduct online A/B test on Kuaishou’s online advertising platform. The traffic of the entire platform was evenly divided into ten buckets. 10\% of the traffic is allocated to the online baseline model, while another 10\% is allocated to HiT-LBM. Our advertising platform serves over 400 million users, and the results collected from 10\% of the traffic over several weeks are highly convincing. We chunk user behaviors that occurred within Kuaishou app (over one year) on a monthly basis. If a chunk contains excessive user behaviors, we prioritize retaining valid samples (e.g., click, conversion) and downsample negative samples (impression only). The user interest representations are then extracted using HiT-LBM. We incorporate these interest representations as side information into the online advertising model for prediction.

\begin{table}[h]
\centering
\renewcommand{\arraystretch}{0.85}
\caption{Results on online advertising platform.}
\label{tab:industrial result}
\begin{tabular}{@{}lccr@{}}
\toprule
Method &  Setting & Revenue & CVR \\ 
\midrule
\multirow{2}{*}{HiT-LBM} & all & \textcolor{red}{$\uparrow$\,3.5\%} & \textcolor{red}{$\uparrow$\,2.3\%} \\
       & long-tail &  \textcolor{red}{$\uparrow$\,4.5\%} & \textcolor{red}{$\uparrow$\,3.1\%} \\
\bottomrule
\end{tabular}
\end{table}

\subsubsection{Experimental Results (RQ3)}
As shown in Table ~\ref{tab:industrial result}, during a 14-day online A/B test, our method achieve an increase in revenue of 3.5\% and an improvement in conversion rate of 2.3\% compared to the baseline, demonstrating significant commercial benefits. Additionally, HiT-LBM is also suitable for users with fewer interactions, addressing the shortcomings of recommendation models in long-tail estimation (i.e., cold start). We validate the effectiveness of handling cold start using data with fewer interactions. Experiments show that HiT-LBM achieved an 4.5\% increase in revenue and an 3.1\% improvement in conversion rate on long-tail data, indicating that the broad reasoning capabilities of LLMs effectively compensate for the limitations of ID-based models in handling long-tail data.

\subsection{Ablation Study}
\begin{figure}[htbp]
\includegraphics[width=0.5\textwidth]{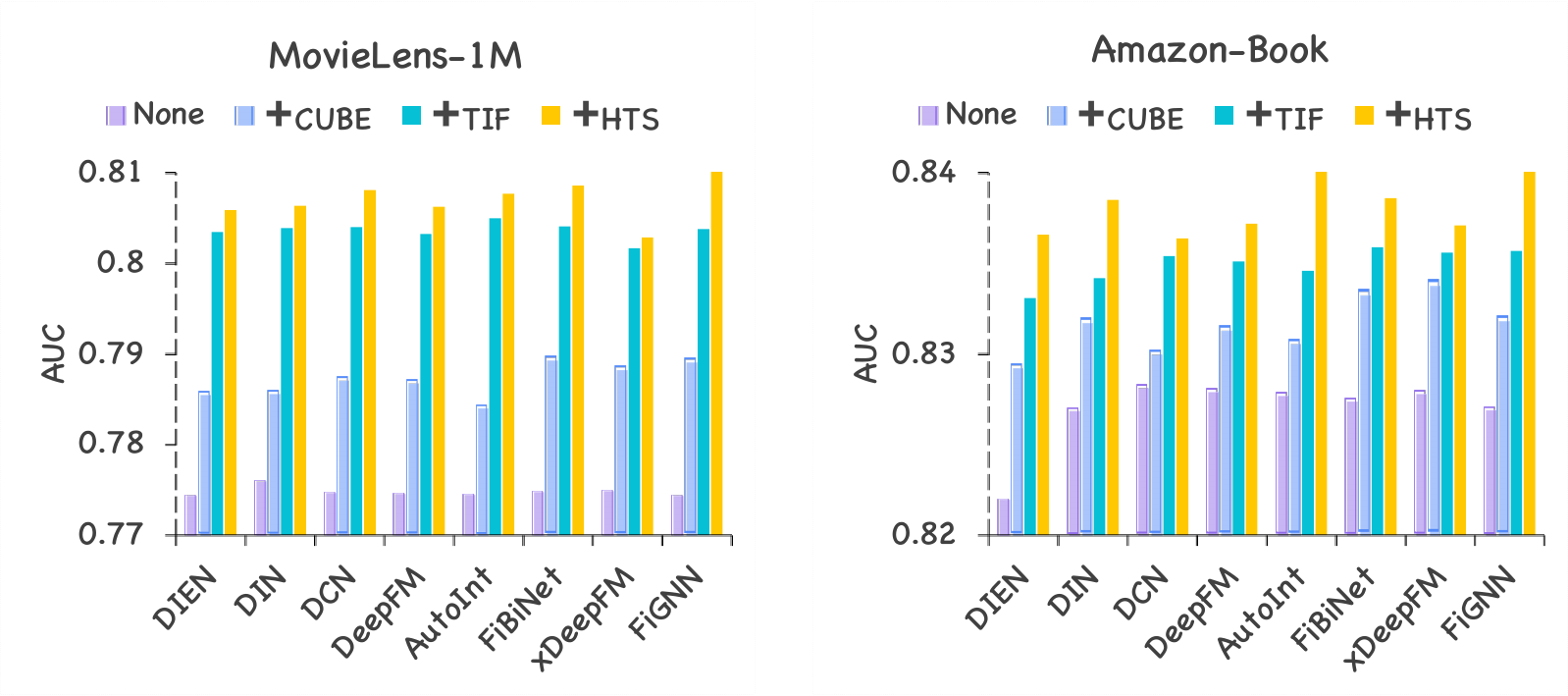}
\caption{The impact of different modules in HiT-LBM on various backbone recommendation models.}
\label{fig:ablation}
\end{figure}

\subsubsection{effect of different modules in HiT-LBM (RQ4).}

In this section, we analyze the impact of different modules of HiT-LBM on recommendation performance. We conduct ablation experiments across different recommendation model backbones. We sequentially introduce chunked user behavior extraction (CUBE), temporal-aware interest fusion(TIF), and hierarchical tree search(HTS). When only CUBE is used, we employ a simple average pooling for interest aggregation. As illustrated in Figure \ref{fig:ablation}, the recommendation performance progressively improves as different backbone networks sequentially incorporate CUBE, TIF, and HTS. We find that CUBE can alleviate the constraints of long sequences on LLMs, thereby enhancing performance. The introduction of TIF significantly boosts recommendation performance, with improvements of 2.28\% on MovieLens-1M and 0.52\% on Amazon-Book, indicating that temporal fusion methods can effectively model the evolution of user interests. Finally, the incorporation of HTS further enhances DIN's recommendation performance, achieving the best recommendation results, which suggests that hierarchical tree search successfully identifies the optimal interest paths for users.

\subsubsection{analysis of the interest fusion module (RQ5).}
\begin{table}[htbp]
\centering
\caption{Performance Comparison of different fusion modules. The best results
are highlighted in bold, while the second-best results are
indicated with underlining.}
\label{tab:fusion}
\begin{tabular}{lcccc}
\toprule
\multirow{2}{*}{\textbf{Module}} & \multicolumn{2}{c}{MovieLens-1M} & \multicolumn{2}{c}{Amazon-Book} \\
\cmidrule(lr){2-3} \cmidrule(lr){4-5} &
\textbf{AUC} & \textbf{Logloss} & \textbf{AUC} & \textbf{Logloss} \\
\midrule
base (DIN) & 0.7761 & 0.5583 & 0.8270& 0.5019\\
MLP & 0.7860 & 0.5491 &  0.8320& 0.4935\\
SA & 0.7963 & \underline{0.5383} & 0.8330 & 0.4929\\
GRU & \underline{0.7978} & 0.5392 & \underline{0.8351} & \underline{0.4919}\\
TIF & \textbf{0.8063}  & \textbf{0.5301} & \textbf{0.8385} & \textbf{0.4886} \\
\bottomrule
\end{tabular}
\end{table}
In this section, we explore the impact of the interest fusion module on traditional recommendation models in the context of interest application. We compare our proposed temporal-aware interest fusion module with currently common fusion methods, using DIN as the backbone network for validation. As shown in Table \ref{tab:fusion}, MLP represents the fusion of user interests through average pooling followed by an MLP, SA and GRU denote the fusion of interests using self-attention mechanisms and Gate Recurrent Unit, respectively, and TIF signifies our temporal-aware interest fusion. We can observe that interest fusion is crucial for enhancing the performance of recommendation models. Furthermore, our proposed TIF achieves the best performance, highlighting the importance of temporal-aware fusion in the representation of user interests.
\subsubsection{effect of different LLM scale (RQ6).}
\begin{figure}[h]
\includegraphics[width=0.3\textwidth]{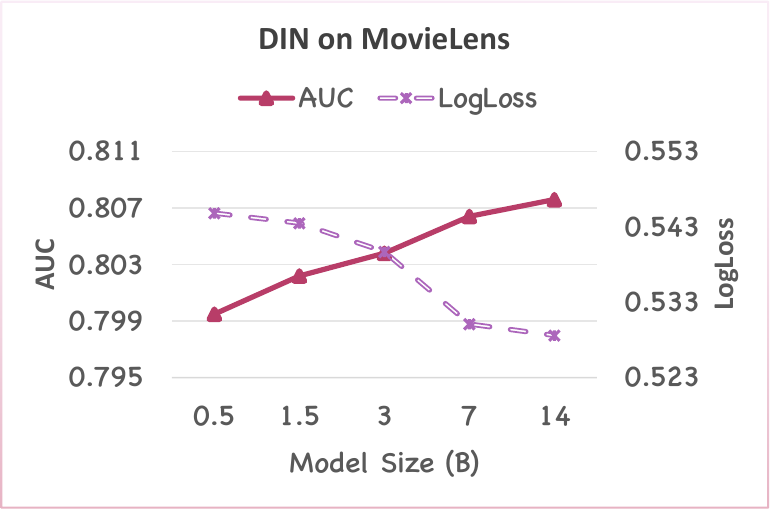}
\caption{The impact of different LLM scale.}
\label{fig:scale}
\end{figure}

To validate the impact of different scales of LLMs on modeling interests, we implement HiT-LBM using various scales of the Qwen2.5 series LLMs. We integrate the interests generated by LLMs of different scales into user representations and embed them into the recommendation model. Figure \ref{fig:scale} illustrates the performance of user interest modeling by LLMs of different scales. We find that the performance of the recommendation model gradually improves as the scale of the LLM increases, indicating that LLMs with larger parameter scales possess stronger capabilities for user interest modeling.
\subsubsection{impact of different interest encoders (RQ7)}
\begin{figure}[h]
\includegraphics[width=0.40\textwidth]{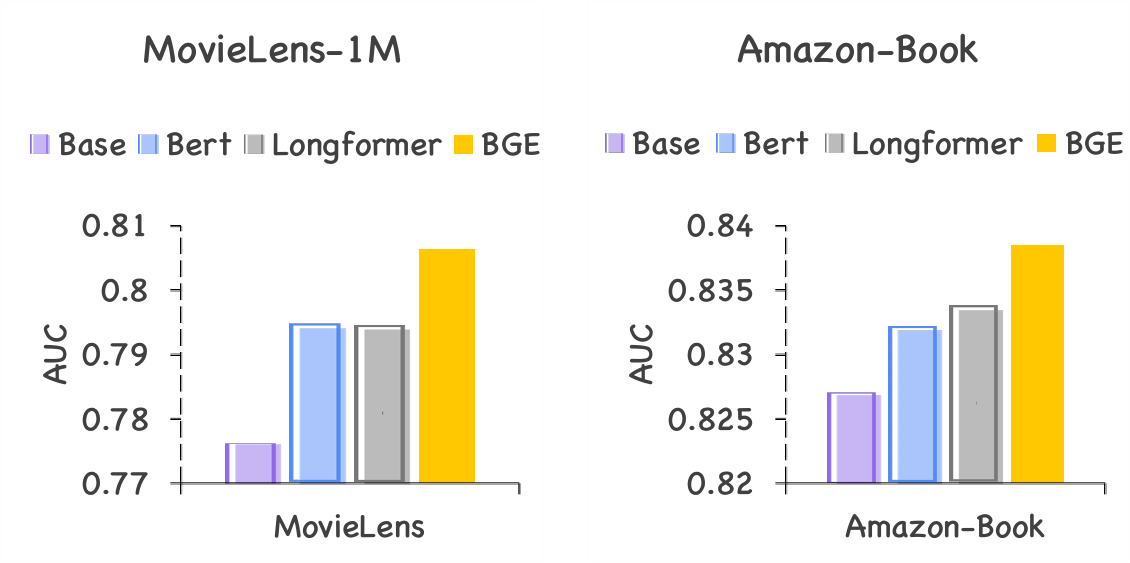}
\caption{The impact of different interest encoder.}
\label{fig:encoder}
\end{figure}
In this section, we investigate the impact of interest encoders during the interest utilization. We employ Bert\cite{bert}, Longformer\cite{longformer}, and BGE\cite{bge} as encoders on the MovieLens-1M and Amazon-Book datasets, respectively, to observe their effects on the performance of the DIN model. As shown in Figure \ref{fig:encoder}, Base represents not using an interest encoder. We find that Longformer and BGE achieve superior performance compared to Bert. This may be due to Bert's limited context processing capability, as it can only handle up to 512 tokens. BGE attains the best performance, indicating that BGE is well-suited for the use of user interests in recommendation models.

\section{Conclusions}
The proposed HiT-LBM framework captures user interest evolution through Chunked User Behavior Extraction (CUBE) and Hierarchical Tree Search (HTS), while constructing lifelong interest representations with Temporal-aware Interest Fusion (TIF). This representation can be embedded into any recommendation model to enhance recommendation performance. Extensive experiments on public and industry datasets demonstrate that this framework is compatible with various ID-based recommendation systems, significantly enhancing their performance. It also effectively addresses the context limitation issues of LLMs in processing long sequences, thereby improving the quality of interest representations.


\bibliographystyle{ACM-Reference-Format}
\balance
\bibliography{main}


\begin{thebibliography}{40}


\ifx \showCODEN    \undefined \def \showCODEN     #1{\unskip}     \fi
\ifx \showISBNx    \undefined \def \showISBNx     #1{\unskip}     \fi
\ifx \showISBNxiii \undefined \def \showISBNxiii  #1{\unskip}     \fi
\ifx \showISSN     \undefined \def \showISSN      #1{\unskip}     \fi
\ifx \showLCCN     \undefined \def \showLCCN      #1{\unskip}     \fi
\ifx \shownote     \undefined \def \shownote      #1{#1}          \fi
\ifx \showarticletitle \undefined \def \showarticletitle #1{#1}   \fi
\ifx \showURL      \undefined \def \showURL       {\relax}        \fi
\providecommand\bibfield[2]{#2}
\providecommand\bibinfo[2]{#2}
\providecommand\natexlab[1]{#1}
\providecommand\showeprint[2][]{arXiv:#2}

\bibitem[Achiam et~al\mbox{.}(2023)]%
        {gpt4}
\bibfield{author}{\bibinfo{person}{Josh Achiam}, \bibinfo{person}{Steven Adler}, \bibinfo{person}{Sandhini Agarwal}, \bibinfo{person}{Lama Ahmad}, \bibinfo{person}{Ilge Akkaya}, \bibinfo{person}{Florencia~Leoni Aleman}, \bibinfo{person}{Diogo Almeida}, \bibinfo{person}{Janko Altenschmidt}, \bibinfo{person}{Sam Altman}, \bibinfo{person}{Shyamal Anadkat}, {et~al\mbox{.}}} \bibinfo{year}{2023}\natexlab{}.
\newblock \showarticletitle{GPT-4 Technical Report}.
\newblock \bibinfo{journal}{\emph{arXiv preprint arXiv:2303.08774}} (\bibinfo{year}{2023}).
\newblock


\bibitem[Azaria et~al\mbox{.}(2013)]%
        {w2}
\bibfield{author}{\bibinfo{person}{Amos Azaria}, \bibinfo{person}{Avinatan Hassidim}, \bibinfo{person}{Sarit Kraus}, \bibinfo{person}{Adi Eshkol}, \bibinfo{person}{Ofer Weintraub}, {and} \bibinfo{person}{Irit Netanely}.} \bibinfo{year}{2013}\natexlab{}.
\newblock \showarticletitle{Movie Recommender System for Profit Maximization}. In \bibinfo{booktitle}{\emph{Proceedings of the 7th ACM Conference on Recommender Systems}}. \bibinfo{pages}{121--128}.
\newblock


\bibitem[Bao et~al\mbox{.}(2023)]%
        {tallrec}
\bibfield{author}{\bibinfo{person}{Keqin Bao}, \bibinfo{person}{Jizhi Zhang}, \bibinfo{person}{Yang Zhang}, \bibinfo{person}{Wenjie Wang}, \bibinfo{person}{Fuli Feng}, {and} \bibinfo{person}{Xiangnan He}.} \bibinfo{year}{2023}\natexlab{}.
\newblock \showarticletitle{Tallrec: An Effective and Efficient Tuning Framework to Align Large Language Model with Recommendation}. In \bibinfo{booktitle}{\emph{Proceedings of the 17th ACM Conference on Recommender Systems}}. \bibinfo{pages}{1007--1014}.
\newblock


\bibitem[Beltagy et~al\mbox{.}(2020)]%
        {longformer}
\bibfield{author}{\bibinfo{person}{Iz Beltagy}, \bibinfo{person}{Matthew~E Peters}, {and} \bibinfo{person}{Arman Cohan}.} \bibinfo{year}{2020}\natexlab{}.
\newblock \showarticletitle{Longformer: The long-document transformer}.
\newblock \bibinfo{journal}{\emph{arXiv preprint arXiv:2004.05150}} (\bibinfo{year}{2020}).
\newblock


\bibitem[Chen et~al\mbox{.}(2024)]%
        {bge}
\bibfield{author}{\bibinfo{person}{Jianlv Chen}, \bibinfo{person}{Shitao Xiao}, \bibinfo{person}{Peitian Zhang}, \bibinfo{person}{Kun Luo}, \bibinfo{person}{Defu Lian}, {and} \bibinfo{person}{Zheng Liu}.} \bibinfo{year}{2024}\natexlab{}.
\newblock \showarticletitle{Bge m3-embedding: Multi-lingual, multi-functionality, multi-granularity text embeddings through self-knowledge distillation}.
\newblock \bibinfo{journal}{\emph{arXiv preprint arXiv:2402.03216}} (\bibinfo{year}{2024}).
\newblock


\bibitem[Devlin(2018)]%
        {bert}
\bibfield{author}{\bibinfo{person}{Jacob Devlin}.} \bibinfo{year}{2018}\natexlab{}.
\newblock \showarticletitle{Bert: Pre-training of deep bidirectional transformers for language understanding}.
\newblock \bibinfo{journal}{\emph{arXiv preprint arXiv:1810.04805}} (\bibinfo{year}{2018}).
\newblock


\bibitem[Dey and Salem(2017)]%
        {gru}
\bibfield{author}{\bibinfo{person}{Rahul Dey} {and} \bibinfo{person}{Fathi~M Salem}.} \bibinfo{year}{2017}\natexlab{}.
\newblock \showarticletitle{Gate-Variants of Gated Recurrent Unit (GRU) Neural Networks}. In \bibinfo{booktitle}{\emph{2017 IEEE 60th International Midwest Symposium on Circuits and Systems (MWSCAS)}}. IEEE, \bibinfo{pages}{1597--1600}.
\newblock


\bibitem[Guo et~al\mbox{.}(2017)]%
        {deepfm}
\bibfield{author}{\bibinfo{person}{Huifeng Guo}, \bibinfo{person}{Ruiming Tang}, \bibinfo{person}{Yunming Ye}, \bibinfo{person}{Zhenguo Li}, {and} \bibinfo{person}{Xiuqiang He}.} \bibinfo{year}{2017}\natexlab{}.
\newblock \showarticletitle{DeepFM: A Factorization-Machine Based Neural Network for CTR Prediction}.
\newblock \bibinfo{journal}{\emph{arXiv preprint arXiv:1703.04247}} (\bibinfo{year}{2017}).
\newblock


\bibitem[He and McAuley(2016)]%
        {w1}
\bibfield{author}{\bibinfo{person}{Ruining He} {and} \bibinfo{person}{Julian McAuley}.} \bibinfo{year}{2016}\natexlab{}.
\newblock \showarticletitle{VBPR: Visual Bayesian Personalized Ranking from Implicit Feedback}. In \bibinfo{booktitle}{\emph{Proceedings of the AAAI Conference on Artificial Intelligence}}, Vol.~\bibinfo{volume}{30}.
\newblock


\bibitem[Ho et~al\mbox{.}(2022)]%
        {cascaded}
\bibfield{author}{\bibinfo{person}{Jonathan Ho}, \bibinfo{person}{Chitwan Saharia}, \bibinfo{person}{William Chan}, \bibinfo{person}{David~J Fleet}, \bibinfo{person}{Mohammad Norouzi}, {and} \bibinfo{person}{Tim Salimans}.} \bibinfo{year}{2022}\natexlab{}.
\newblock \showarticletitle{Cascaded Diffusion Models for High Fidelity Image Generation}.
\newblock \bibinfo{journal}{\emph{Journal of Machine Learning Research}} \bibinfo{volume}{23}, \bibinfo{number}{47} (\bibinfo{year}{2022}), \bibinfo{pages}{1--33}.
\newblock


\bibitem[Hu et~al\mbox{.}(2024)]%
        {hu2024can}
\bibfield{author}{\bibinfo{person}{Yutong Hu}, \bibinfo{person}{Quzhe Huang}, \bibinfo{person}{Mingxu Tao}, \bibinfo{person}{Chen Zhang}, {and} \bibinfo{person}{Yansong Feng}.} \bibinfo{year}{2024}\natexlab{}.
\newblock \showarticletitle{Can Perplexity Reflect Large Language Model's Ability in Long Text Understanding?}
\newblock \bibinfo{journal}{\emph{arXiv preprint arXiv:2405.06105}} (\bibinfo{year}{2024}).
\newblock


\bibitem[Huang et~al\mbox{.}(2019)]%
        {fibinet}
\bibfield{author}{\bibinfo{person}{Tongwen Huang}, \bibinfo{person}{Zhiqi Zhang}, {and} \bibinfo{person}{Junlin Zhang}.} \bibinfo{year}{2019}\natexlab{}.
\newblock \showarticletitle{FiBiNET: Combining Feature Importance and Bilinear Feature Interaction for Click-Through Rate Prediction}. In \bibinfo{booktitle}{\emph{Proceedings of the 13th ACM Conference on Recommender Systems}}. \bibinfo{pages}{169--177}.
\newblock


\bibitem[Li et~al\mbox{.}(2023)]%
        {li2023loogle}
\bibfield{author}{\bibinfo{person}{Jiaqi Li}, \bibinfo{person}{Mengmeng Wang}, \bibinfo{person}{Zilong Zheng}, {and} \bibinfo{person}{Muhan Zhang}.} \bibinfo{year}{2023}\natexlab{}.
\newblock \showarticletitle{LooGLE: Can Long-Context Language Models Understand Long Contexts?}
\newblock \bibinfo{journal}{\emph{arXiv preprint arXiv:2311.04939}} (\bibinfo{year}{2023}).
\newblock


\bibitem[Li et~al\mbox{.}(2019)]%
        {fignn}
\bibfield{author}{\bibinfo{person}{Zekun Li}, \bibinfo{person}{Zeyu Cui}, \bibinfo{person}{Shu Wu}, \bibinfo{person}{Xiaoyu Zhang}, {and} \bibinfo{person}{Liang Wang}.} \bibinfo{year}{2019}\natexlab{}.
\newblock \showarticletitle{Fi-gnn: Modeling feature interactions via graph neural networks for ctr prediction}. In \bibinfo{booktitle}{\emph{Proceedings of the 28th ACM international conference on information and knowledge management}}. \bibinfo{pages}{539--548}.
\newblock


\bibitem[Lian et~al\mbox{.}(2018)]%
        {xdeepfm}
\bibfield{author}{\bibinfo{person}{Jianxun Lian}, \bibinfo{person}{Xiaohuan Zhou}, \bibinfo{person}{Fuzheng Zhang}, \bibinfo{person}{Zhongxia Chen}, \bibinfo{person}{Xing Xie}, {and} \bibinfo{person}{Guangzhong Sun}.} \bibinfo{year}{2018}\natexlab{}.
\newblock \showarticletitle{xDeepFM: Combining Explicit and Implicit Feature Interactions for Recommender Systems}. In \bibinfo{booktitle}{\emph{Proceedings of the 24th ACM SIGKDD International Conference on Knowledge Discovery \& Data Mining}}. \bibinfo{pages}{1754--1763}.
\newblock


\bibitem[Lin et~al\mbox{.}(2024)]%
        {rella}
\bibfield{author}{\bibinfo{person}{Jianghao Lin}, \bibinfo{person}{Rong Shan}, \bibinfo{person}{Chenxu Zhu}, \bibinfo{person}{Kounianhua Du}, \bibinfo{person}{Bo Chen}, \bibinfo{person}{Shigang Quan}, \bibinfo{person}{Ruiming Tang}, \bibinfo{person}{Yong Yu}, {and} \bibinfo{person}{Weinan Zhang}.} \bibinfo{year}{2024}\natexlab{}.
\newblock \showarticletitle{Rella: Retrieval-Enhanced Large Language Models for Lifelong Sequential Behavior Comprehension in Recommendation}. In \bibinfo{booktitle}{\emph{Proceedings of the ACM on Web Conference 2024}}. \bibinfo{pages}{3497--3508}.
\newblock


\bibitem[Liu et~al\mbox{.}(2024a)]%
        {deepseek}
\bibfield{author}{\bibinfo{person}{Aixin Liu}, \bibinfo{person}{Bei Feng}, \bibinfo{person}{Bing Xue}, \bibinfo{person}{Bingxuan Wang}, \bibinfo{person}{Bochao Wu}, \bibinfo{person}{Chengda Lu}, \bibinfo{person}{Chenggang Zhao}, \bibinfo{person}{Chengqi Deng}, \bibinfo{person}{Chenyu Zhang}, \bibinfo{person}{Chong Ruan}, {et~al\mbox{.}}} \bibinfo{year}{2024}\natexlab{a}.
\newblock \showarticletitle{Deepseek-v3 Technical Report}.
\newblock \bibinfo{journal}{\emph{arXiv preprint arXiv:2412.19437}} (\bibinfo{year}{2024}).
\newblock


\bibitem[Liu et~al\mbox{.}(2024b)]%
        {mamba4rec}
\bibfield{author}{\bibinfo{person}{Chengkai Liu}, \bibinfo{person}{Jianghao Lin}, \bibinfo{person}{Jianling Wang}, \bibinfo{person}{Hanzhou Liu}, {and} \bibinfo{person}{James Caverlee}.} \bibinfo{year}{2024}\natexlab{b}.
\newblock \showarticletitle{Mamba4rec: Towards Efficient Sequential Recommendation with Selective State Space Models}.
\newblock \bibinfo{journal}{\emph{arXiv preprint arXiv:2403.03900}} (\bibinfo{year}{2024}).
\newblock


\bibitem[Pi et~al\mbox{.}(2020)]%
        {sim}
\bibfield{author}{\bibinfo{person}{Qi Pi}, \bibinfo{person}{Guorui Zhou}, \bibinfo{person}{Yujing Zhang}, \bibinfo{person}{Zhe Wang}, \bibinfo{person}{Lejian Ren}, \bibinfo{person}{Ying Fan}, \bibinfo{person}{Xiaoqiang Zhu}, {and} \bibinfo{person}{Kun Gai}.} \bibinfo{year}{2020}\natexlab{}.
\newblock \showarticletitle{Search-Based User Interest Modeling with Lifelong Sequential Behavior Data for Click-Through Rate Prediction}. In \bibinfo{booktitle}{\emph{Proceedings of the 29th ACM International Conference on Information \& Knowledge Management}}. \bibinfo{pages}{2685--2692}.
\newblock


\bibitem[Ravaut et~al\mbox{.}(2024)]%
        {context}
\bibfield{author}{\bibinfo{person}{Mathieu Ravaut}, \bibinfo{person}{Aixin Sun}, \bibinfo{person}{Nancy Chen}, {and} \bibinfo{person}{Shafiq Joty}.} \bibinfo{year}{2024}\natexlab{}.
\newblock \showarticletitle{On Context Utilization in Summarization with Large Language Models}. In \bibinfo{booktitle}{\emph{Proceedings of the 62nd Annual Meeting of the Association for Computational Linguistics (Volume 1: Long Papers)}}. \bibinfo{pages}{2764--2781}.
\newblock


\bibitem[Ren et~al\mbox{.}(2024)]%
        {llmrec_3}
\bibfield{author}{\bibinfo{person}{Yankun Ren}, \bibinfo{person}{Zhongde Chen}, \bibinfo{person}{Xinxing Yang}, \bibinfo{person}{Longfei Li}, \bibinfo{person}{Cong Jiang}, \bibinfo{person}{Lei Cheng}, \bibinfo{person}{Bo Zhang}, \bibinfo{person}{Linjian Mo}, {and} \bibinfo{person}{Jun Zhou}.} \bibinfo{year}{2024}\natexlab{}.
\newblock \showarticletitle{Enhancing Sequential Recommenders with Augmented Knowledge from Aligned Large Language Models}. In \bibinfo{booktitle}{\emph{Proceedings of the 47th International {ACM} {SIGIR} Conference on Research and Development in Information Retrieval, {SIGIR} 2024, Washington DC, USA, July 14-18, 2024}}, \bibfield{editor}{\bibinfo{person}{Grace~Hui Yang}, \bibinfo{person}{Hongning Wang}, \bibinfo{person}{Sam Han}, \bibinfo{person}{Claudia Hauff}, \bibinfo{person}{Guido Zuccon}, {and} \bibinfo{person}{Yi~Zhang}} (Eds.). \bibinfo{publisher}{{ACM}}, \bibinfo{pages}{345--354}.
\newblock
\href{https://doi.org/10.1145/3626772.3657782}{doi:\nolinkurl{10.1145/3626772.3657782}}


\bibitem[Song et~al\mbox{.}(2019)]%
        {autoint}
\bibfield{author}{\bibinfo{person}{Weiping Song}, \bibinfo{person}{Chence Shi}, \bibinfo{person}{Zhiping Xiao}, \bibinfo{person}{Zhijian Duan}, \bibinfo{person}{Yewen Xu}, \bibinfo{person}{Ming Zhang}, {and} \bibinfo{person}{Jian Tang}.} \bibinfo{year}{2019}\natexlab{}.
\newblock \showarticletitle{AutoInt: Automatic Feature Interaction Learning via Self-Attentive Neural Networks}. In \bibinfo{booktitle}{\emph{Proceedings of the 28th ACM International Conference on Information and Knowledge Management}}. \bibinfo{pages}{1161--1170}.
\newblock


\bibitem[Sun et~al\mbox{.}(2019)]%
        {bert4rec}
\bibfield{author}{\bibinfo{person}{Fei Sun}, \bibinfo{person}{Jun Liu}, \bibinfo{person}{Jian Wu}, \bibinfo{person}{Changhua Pei}, \bibinfo{person}{Xiao Lin}, \bibinfo{person}{Wenwu Ou}, {and} \bibinfo{person}{Peng Jiang}.} \bibinfo{year}{2019}\natexlab{}.
\newblock \showarticletitle{BERT4Rec: Sequential Recommendation with Bidirectional Encoder Representations from Transformer}. In \bibinfo{booktitle}{\emph{Proceedings of the 28th ACM International Conference on Information and Knowledge Management}}. \bibinfo{pages}{1441--1450}.
\newblock


\bibitem[Sun et~al\mbox{.}(2024)]%
        {llmrec_1}
\bibfield{author}{\bibinfo{person}{Zhu Sun}, \bibinfo{person}{Hongyang Liu}, \bibinfo{person}{Xinghua Qu}, \bibinfo{person}{Kaidong Feng}, \bibinfo{person}{Yan Wang}, {and} \bibinfo{person}{Yew~Soon Ong}.} \bibinfo{year}{2024}\natexlab{}.
\newblock \showarticletitle{Large Language Models for Intent-Driven Session Recommendations}. In \bibinfo{booktitle}{\emph{Proceedings of the 47th International {ACM} {SIGIR} Conference on Research and Development in Information Retrieval, {SIGIR} 2024, Washington DC, USA, July 14-18, 2024}}, \bibfield{editor}{\bibinfo{person}{Grace~Hui Yang}, \bibinfo{person}{Hongning Wang}, \bibinfo{person}{Sam Han}, \bibinfo{person}{Claudia Hauff}, \bibinfo{person}{Guido Zuccon}, {and} \bibinfo{person}{Yi~Zhang}} (Eds.). \bibinfo{publisher}{{ACM}}, \bibinfo{pages}{324--334}.
\newblock
\href{https://doi.org/10.1145/3626772.3657688}{doi:\nolinkurl{10.1145/3626772.3657688}}


\bibitem[Touvron et~al\mbox{.}(2023)]%
        {llama}
\bibfield{author}{\bibinfo{person}{Hugo Touvron}, \bibinfo{person}{Thibaut Lavril}, \bibinfo{person}{Gautier Izacard}, \bibinfo{person}{Xavier Martinet}, \bibinfo{person}{Marie-Anne Lachaux}, \bibinfo{person}{Timoth{\'e}e Lacroix}, \bibinfo{person}{Baptiste Rozi{\`e}re}, \bibinfo{person}{Naman Goyal}, \bibinfo{person}{Eric Hambro}, \bibinfo{person}{Faisal Azhar}, {et~al\mbox{.}}} \bibinfo{year}{2023}\natexlab{}.
\newblock \showarticletitle{Llama: Open and Efficient Foundation Language Models}.
\newblock \bibinfo{journal}{\emph{arXiv preprint arXiv:2302.13971}} (\bibinfo{year}{2023}).
\newblock


\bibitem[Van~den Oord et~al\mbox{.}(2013)]%
        {w3}
\bibfield{author}{\bibinfo{person}{Aaron Van~den Oord}, \bibinfo{person}{Sander Dieleman}, {and} \bibinfo{person}{Benjamin Schrauwen}.} \bibinfo{year}{2013}\natexlab{}.
\newblock \showarticletitle{Deep Content-Based Music Recommendation}.
\newblock \bibinfo{journal}{\emph{Advances in Neural Information Processing Systems}}  \bibinfo{volume}{26} (\bibinfo{year}{2013}).
\newblock


\bibitem[Vaswani(2017)]%
        {attention}
\bibfield{author}{\bibinfo{person}{A Vaswani}.} \bibinfo{year}{2017}\natexlab{}.
\newblock \showarticletitle{Attention is All You Need}.
\newblock \bibinfo{journal}{\emph{Advances in Neural Information Processing Systems}} (\bibinfo{year}{2017}).
\newblock


\bibitem[Wang et~al\mbox{.}(2024)]%
        {llmrec_4}
\bibfield{author}{\bibinfo{person}{Jie Wang}, \bibinfo{person}{Alexandros Karatzoglou}, \bibinfo{person}{Ioannis Arapakis}, {and} \bibinfo{person}{Joemon~M. Jose}.} \bibinfo{year}{2024}\natexlab{}.
\newblock \showarticletitle{Reinforcement Learning-based Recommender Systems with Large Language Models for State Reward and Action Modeling}. In \bibinfo{booktitle}{\emph{Proceedings of the 47th International {ACM} {SIGIR} Conference on Research and Development in Information Retrieval, {SIGIR} 2024, Washington DC, USA, July 14-18, 2024}}, \bibfield{editor}{\bibinfo{person}{Grace~Hui Yang}, \bibinfo{person}{Hongning Wang}, \bibinfo{person}{Sam Han}, \bibinfo{person}{Claudia Hauff}, \bibinfo{person}{Guido Zuccon}, {and} \bibinfo{person}{Yi~Zhang}} (Eds.). \bibinfo{publisher}{{ACM}}, \bibinfo{pages}{375--385}.
\newblock
\href{https://doi.org/10.1145/3626772.3657767}{doi:\nolinkurl{10.1145/3626772.3657767}}


\bibitem[Wang et~al\mbox{.}(2017)]%
        {dcnv1}
\bibfield{author}{\bibinfo{person}{Ruoxi Wang}, \bibinfo{person}{Bin Fu}, \bibinfo{person}{Gang Fu}, {and} \bibinfo{person}{Mingliang Wang}.} \bibinfo{year}{2017}\natexlab{}.
\newblock \showarticletitle{Deep \& Cross Network for Ad Click Predictions}.
\newblock In \bibinfo{booktitle}{\emph{Proceedings of the ADKDD'17}}. \bibinfo{pages}{1--7}.
\newblock


\bibitem[Wang et~al\mbox{.}(2021)]%
        {dcn}
\bibfield{author}{\bibinfo{person}{Ruoxi Wang}, \bibinfo{person}{Rakesh Shivanna}, \bibinfo{person}{Derek Cheng}, \bibinfo{person}{Sagar Jain}, \bibinfo{person}{Dong Lin}, \bibinfo{person}{Lichan Hong}, {and} \bibinfo{person}{Ed Chi}.} \bibinfo{year}{2021}\natexlab{}.
\newblock \showarticletitle{DCN V2: Improved Deep \& Cross Network and Practical Lessons for Web-Scale Learning to Rank Systems}. In \bibinfo{booktitle}{\emph{Proceedings of the Web Conference 2021}}. \bibinfo{pages}{1785--1797}.
\newblock


\bibitem[Wei et~al\mbox{.}(2024)]%
        {llmrec}
\bibfield{author}{\bibinfo{person}{Wei Wei}, \bibinfo{person}{Xubin Ren}, \bibinfo{person}{Jiabin Tang}, \bibinfo{person}{Qinyong Wang}, \bibinfo{person}{Lixin Su}, \bibinfo{person}{Suqi Cheng}, \bibinfo{person}{Junfeng Wang}, \bibinfo{person}{Dawei Yin}, {and} \bibinfo{person}{Chao Huang}.} \bibinfo{year}{2024}\natexlab{}.
\newblock \showarticletitle{Llmrec: Large Language Models with Graph Augmentation for Recommendation}. In \bibinfo{booktitle}{\emph{Proceedings of the 17th ACM International Conference on Web Search and Data Mining}}. \bibinfo{pages}{806--815}.
\newblock


\bibitem[Xi et~al\mbox{.}(2024)]%
        {kar}
\bibfield{author}{\bibinfo{person}{Yunjia Xi}, \bibinfo{person}{Weiwen Liu}, \bibinfo{person}{Jianghao Lin}, \bibinfo{person}{Xiaoling Cai}, \bibinfo{person}{Hong Zhu}, \bibinfo{person}{Jieming Zhu}, \bibinfo{person}{Bo Chen}, \bibinfo{person}{Ruiming Tang}, \bibinfo{person}{Weinan Zhang}, {and} \bibinfo{person}{Yong Yu}.} \bibinfo{year}{2024}\natexlab{}.
\newblock \showarticletitle{Towards Open-World Recommendation with Knowledge Augmentation from Large Language Models}. In \bibinfo{booktitle}{\emph{Proceedings of the 18th ACM Conference on Recommender Systems}}. \bibinfo{pages}{12--22}.
\newblock


\bibitem[Yang et~al\mbox{.}(2024b)]%
        {qwen}
\bibfield{author}{\bibinfo{person}{An Yang}, \bibinfo{person}{Baosong Yang}, \bibinfo{person}{Beichen Zhang}, \bibinfo{person}{Binyuan Hui}, \bibinfo{person}{Bo Zheng}, \bibinfo{person}{Bowen Yu}, \bibinfo{person}{Chengyuan Li}, \bibinfo{person}{Dayiheng Liu}, \bibinfo{person}{Fei Huang}, \bibinfo{person}{Haoran Wei}, {et~al\mbox{.}}} \bibinfo{year}{2024}\natexlab{b}.
\newblock \showarticletitle{Qwen2.5 Technical Report}.
\newblock \bibinfo{journal}{\emph{arXiv preprint arXiv:2412.15115}} (\bibinfo{year}{2024}).
\newblock


\bibitem[Yang et~al\mbox{.}(2024a)]%
        {llmrec_2}
\bibfield{author}{\bibinfo{person}{Shenghao Yang}, \bibinfo{person}{Weizhi Ma}, \bibinfo{person}{Peijie Sun}, \bibinfo{person}{Qingyao Ai}, \bibinfo{person}{Yiqun Liu}, \bibinfo{person}{Mingchen Cai}, {and} \bibinfo{person}{Min Zhang}.} \bibinfo{year}{2024}\natexlab{a}.
\newblock \showarticletitle{Sequential Recommendation with Latent Relations based on Large Language Model}. In \bibinfo{booktitle}{\emph{Proceedings of the 47th International {ACM} {SIGIR} Conference on Research and Development in Information Retrieval, {SIGIR} 2024, Washington DC, USA, July 14-18, 2024}}, \bibfield{editor}{\bibinfo{person}{Grace~Hui Yang}, \bibinfo{person}{Hongning Wang}, \bibinfo{person}{Sam Han}, \bibinfo{person}{Claudia Hauff}, \bibinfo{person}{Guido Zuccon}, {and} \bibinfo{person}{Yi~Zhang}} (Eds.). \bibinfo{publisher}{{ACM}}, \bibinfo{pages}{335--344}.
\newblock
\href{https://doi.org/10.1145/3626772.3657762}{doi:\nolinkurl{10.1145/3626772.3657762}}


\bibitem[Zhang et~al\mbox{.}(2024)]%
        {binllm}
\bibfield{author}{\bibinfo{person}{Yang Zhang}, \bibinfo{person}{Keqin Bao}, \bibinfo{person}{Ming Yan}, \bibinfo{person}{Wenjie Wang}, \bibinfo{person}{Fuli Feng}, {and} \bibinfo{person}{Xiangnan He}.} \bibinfo{year}{2024}\natexlab{}.
\newblock \showarticletitle{Text-Like Encoding of Collaborative Information in Large Language Models for Recommendation}.
\newblock \bibinfo{journal}{\emph{arXiv preprint arXiv:2406.03210}} (\bibinfo{year}{2024}).
\newblock


\bibitem[Zhang et~al\mbox{.}(2023)]%
        {collm}
\bibfield{author}{\bibinfo{person}{Yang Zhang}, \bibinfo{person}{Fuli Feng}, \bibinfo{person}{Jizhi Zhang}, \bibinfo{person}{Keqin Bao}, \bibinfo{person}{Qifan Wang}, {and} \bibinfo{person}{Xiangnan He}.} \bibinfo{year}{2023}\natexlab{}.
\newblock \showarticletitle{Collm: Integrating Collaborative Embeddings into Large Language Models for Recommendation}.
\newblock \bibinfo{journal}{\emph{arXiv preprint arXiv:2310.19488}} (\bibinfo{year}{2023}).
\newblock


\bibitem[Zheng et~al\mbox{.}(2024)]%
        {trsr}
\bibfield{author}{\bibinfo{person}{Zhi Zheng}, \bibinfo{person}{Wenshuo Chao}, \bibinfo{person}{Zhaopeng Qiu}, \bibinfo{person}{Hengshu Zhu}, {and} \bibinfo{person}{Hui Xiong}.} \bibinfo{year}{2024}\natexlab{}.
\newblock \showarticletitle{Harnessing large language models for text-rich sequential recommendation}. In \bibinfo{booktitle}{\emph{Proceedings of the ACM on Web Conference 2024}}. \bibinfo{pages}{3207--3216}.
\newblock


\bibitem[Zhou et~al\mbox{.}(2019)]%
        {dien}
\bibfield{author}{\bibinfo{person}{Guorui Zhou}, \bibinfo{person}{Na Mou}, \bibinfo{person}{Ying Fan}, \bibinfo{person}{Qi Pi}, \bibinfo{person}{Weijie Bian}, \bibinfo{person}{Chang Zhou}, \bibinfo{person}{Xiaoqiang Zhu}, {and} \bibinfo{person}{Kun Gai}.} \bibinfo{year}{2019}\natexlab{}.
\newblock \showarticletitle{Deep Interest Evolution Network for Click-Through Rate Prediction}. In \bibinfo{booktitle}{\emph{Proceedings of the AAAI Conference on Artificial Intelligence}}, Vol.~\bibinfo{volume}{33}. \bibinfo{pages}{5941--5948}.
\newblock


\bibitem[Zhou et~al\mbox{.}(2018)]%
        {din}
\bibfield{author}{\bibinfo{person}{Guorui Zhou}, \bibinfo{person}{Xiaoqiang Zhu}, \bibinfo{person}{Chenru Song}, \bibinfo{person}{Ying Fan}, \bibinfo{person}{Han Zhu}, \bibinfo{person}{Xiao Ma}, \bibinfo{person}{Yanghui Yan}, \bibinfo{person}{Junqi Jin}, \bibinfo{person}{Han Li}, {and} \bibinfo{person}{Kun Gai}.} \bibinfo{year}{2018}\natexlab{}.
\newblock \showarticletitle{Deep Interest Network for Click-Through Rate Prediction}. In \bibinfo{booktitle}{\emph{Proceedings of the 24th ACM SIGKDD International Conference on Knowledge Discovery \& Data Mining}}. \bibinfo{pages}{1059--1068}.
\newblock


\bibitem[Zhu et~al\mbox{.}(2024)]%
        {liber}
\bibfield{author}{\bibinfo{person}{Chenxu Zhu}, \bibinfo{person}{Shigang Quan}, \bibinfo{person}{Bo Chen}, \bibinfo{person}{Jianghao Lin}, \bibinfo{person}{Xiaoling Cai}, \bibinfo{person}{Hong Zhu}, \bibinfo{person}{Xiangyang Li}, \bibinfo{person}{Yunjia Xi}, \bibinfo{person}{Weinan Zhang}, {and} \bibinfo{person}{Ruiming Tang}.} \bibinfo{year}{2024}\natexlab{}.
\newblock \showarticletitle{LIBER: Lifelong User Behavior Modeling Based on Large Language Models}.
\newblock \bibinfo{journal}{\emph{arXiv preprint arXiv:2411.14713}} (\bibinfo{year}{2024}).
\newblock


\end{thebibliography}

\appendix









\end{document}